\documentclass[letterpaper]{article}

\usepackage{natbib,alifeconf}  
\usepackage{booktabs}
\usepackage{xcolor}
\usepackage{amsmath}
\usepackage{amssymb}
\usepackage{dsfont}
\usepackage[colorlinks=true, linkcolor=blue, urlcolor=blue, citecolor=blue]{hyperref}
\usepackage{url,cleveref}
\usepackage{booktabs}
\usepackage{subcaption}
\usepackage{nameref}

\renewcommand{\vec}[1]{\mathbf{#1}}
\usepackage{environ}
\NewEnviron{equations}{%
  \begin{equation}
  \begin{split}
    \BODY
  \end{split}
  \end{equation}
}

\title{MaCE: General Mass Conserving Dynamics for Cellular Automata}
\author {Vassilis Papadopoulos$^{1}$\thanks{\,\,\,Equal contributions.} { }\thanks{\,\,\,Corresponding author: vassilis.physics@gmail.com} \and Etienne Guichard$^{2}$\protect\footnotemark[1]
\mbox{}\\
$^1$ CSFT, EPFL, Switzerland\\
$^2$ Independent, Delft, Netherlands
}

\begin{document}

\maketitle
\begin{abstract}
We present Mass-Conserving Evolution (MaCE), a general method for implementing mass conservation in Cellular Automata (CA). MaCE is a simple evolution rule that can be easily 'attached' to existing CAs to make them mass-conserving, which tends to produce interesting behaviours more often, as patterns can no longer explode or die out. We first show that MaCE is numerically stable and admits a simple continuous limit. We then test MaCE on Lenia, and through several experiments, we demonstrate that it produces a wide variety of interesting behaviours, starting from the variety and abundance of solitons up to hints of intrinsic evolution in resource-constrained environments. Finally, we showcase the versatility of MaCE by applying it to Neural-CAs and discrete CAs, and discuss promising research directions opened up by this scheme.
\end{abstract}


Code available at \href{https://github.com/frotaur/MaceLenia}{this link}. Videos available at \href{https://etimush.github.io/MaCE-Videos/}{this link}

\section{Introduction}
In the arduous search for Artificial Life, researchers over the years have produced a great diversity of systems that attempt to capture some of the elusive aspects of life, such as emergence of complexity, self-organization or evolution. Among these models, one particularly well-represented class is that of Cellular Automata (CA)(\cite{vonneumann1966theory,wolframUniversalityComplexityCellular1984}), which come in a host of variations, be it discrete, as the famous Game Of Life (\cite{MathematicalGamesFantastic1970}), continuous, as in the more modern Lenia (\cite{chanLeniaExpandedUniverse2020}) or somewhere in between like for Multi-Neighborhood-CAs (\cite{evansLargerLifeDigital2001,slackermanzMNCACompilationSlackermanz2023}) or Neural-CAs (\cite{mordvintsevGrowingNeuralCellular2020}). 

The appeal of CAs is two-fold. First is their simplicity: defining their behaviours only requires specifying what happens to a cell given its neighborhood, and this simplicity does not preclude them from displaying arbitrarily complex behaviours (\cite{cookUniversalityElementaryCellular2004,johnston2022conways})! Secondly, their inherently local evolution rule means the dynamics are easily parallelizable, allowing the efficient simulation of large-scale worlds necessary to probe emergent properties.

Despite these advantages, one property often found lacking in CAs is the presence of any conserved quantities, most notably, the absence of mass conservation. Although it is by no means necessary to obtain interesting behaviours, it does bring a host of desirable properties. First, the existence of conserved quantities immediately implies limited resources, which are an important factor in driving intrinsic evolution by forcing emerging creatures to compete for access to these resources(\cite{hickinbothamConservationMatterIncreases2015}). Second, it has a strong regularizing behaviour; patterns can no longer explode or die out, and this makes it much more likely for interesting behaviours to arise, as one no longer needs to first look for self-stabilizing sets of rules. Finally, it also allows one to set up interesting experiments, for example by adding extra mass to the system in the form of 'food' or adding mass sinks and sources to break the symmetry of the world. 

Mass conservation in Alife systems is so far mostly achieved by using particle systems, such as Particle Life(\cite{mohrTommohrParticlelifeapp2025}), Particle Lenia (\cite{mordvintsevParticleLeniaEnergybased}), A.L.I.E.N. (\cite{heinemannChrxhAlien2025}) and others (\cite{succi2001lattice,reynoldsFlocksHerdsSchools1987}). While this provides all of the aforementioned advantages, it has other drawbacks. The system is inherently discrete, and approaching a continuous limit involves hugely scaling the number of particles, which becomes an involved endeavour since the complexity of the simulation grows quadratically with the particle number, though this can be optimized up to a point. Compared to CAs, particle systems are mostly an orthogonal approach to Artificial Life, with some trade-offs in simplicity and native parallelizability.

To bring the benefits of mass conservation to CAs, we propose in this work the \textbf{Ma}ss \textbf{C}onserving \textbf{E}volution rule (MaCE), which is an update rule that implements mass conservation locally, and that can be easily slotted into existing CAs (both continuous and discrete) often with little to no changes. The approach taken is similar in spirit to the mass-conserving scheme used in Flow-Lenia(\cite{plantecFlowLeniaOpenendedEvolution2023a}), although we argue our scheme is simpler, while having a wider applicability.

The paper is organized as follows. We begin by introducing MaCE, explaining the logical reasoning behind it, and show it admits a well-defined continuous limit, as well as a version applicable to CAs with discrete states. Then, we probe its capabilities by applying it to Lenia to obtain MaCELenia, which also allows a one-to-one comparison with Flow-Lenia. Finally, we show its versatility by applying it to NCAs and discrete holistic CAs, before closing with a discussion on the possible future applications of MaCE, discussing both its strengths and its shortcomings.

\section{The Mass Conserving Evolution Rule}
In this section, we describe the MaCE rule in detail, prove it conserves mass exactly, and investigate its continuous limit.
\subsection{Definitions}
We denote the 'mass' field as $\rho^t_c(\vec{x})\in \mathbb{R}$, where $t \in \mathbb{R}$ denotes the time, $\vec{x}\in \mathbb{R}^n$ denotes the spatial position (we take $n=2$ from here on) and $c\in [[1,...,C]]$ denotes the different channels. Most of the time we omit the channel indices for readability, and each equation should be thought as acting on each channel. We will also use the discretized versions of these quantities when describing the implementation of the evolution rule, namely $\rho^t_{ij}= \rho^t(i*\Delta x,j*\Delta y)$ where $i,j \in \mathbb{Z}$ denote locations on a grid, and the time $t\in \mathbb{N}$ takes discrete values.

We will also need an 'Affinity' field $A(\vec{x},\rho) \in \mathbb{R}$ ($A_{ij}$ in its discrete version, omitting $\rho$) which is a number that can be computed at each location from the current state of the automaton (the naming is borrowed from Flow-Lenia (\cite{plantecFlowLeniaOpenendedEvolution2023a})). Roughly, this affinity value will determine how much a location is locally 'attractive' to mass; according to the evolution rule, mass will tend to move towards higher affinity values. The specific way in which this affinity value is computed is unconstrained, and as such, any model that computes a real value for each cell is amenable to this mass-conserving modification.

\subsection{Mass-Conserving Evolution Rule}
Given the affinity field, under one time-step the mass field will be evolved according to the following evolution equation
\footnote{These equations are highly inspired by (\cite{ereb0slabs2022}) and we attempted to contact the author without success.}

\begin{align}
Z_{ij} &= \sum_{k,l\in\mathcal{N}_{ij}}e^{\beta A_{kl}}\label{eq:Zcomp}\\
\rho^{t+1}_{ij}&=\sum_{i',j'\in \mathcal{N}_{ij}} \frac{e^{\beta A_{ij}}}{Z_{i'j'}}\rho^t_{i'j'}\label{eq:rhocomp}
\end{align}
Where the sum $k,l\in \mathcal{N}_{ij}$ denotes summing over the neighborhood of cell $(i,j)$. We use a $3\times 3$ Moore neighborhood, so $\mathcal{N}$ will henceforth denote this neighborhood. We also introduced a new parameter, $\beta \in \mathbb{R}$, which we will see can be interpreted as an inverse temperature. 

Let us unpack this evolution equation to understand it. First, instead of working directly with the affinities, we exponentiate them, as we will need positive values. Then, each cell computes $Z_{ij}$, which is a normalization constant for the affinities of its neighbours. The quantity $p_{i'j'} = \frac{e^{\beta A_{i'j'}}}{Z_{ij}}, \,\, (i',j')\in \mathcal{N}_{ij}$ can now be seen as the mass 'portion' each cell in $\mathcal{N}_{ij}$ is entitled to. The higher the relative affinity of a cell compared to the neighborhood, the bigger the mass portion it will receive. The evolution rule then redistributes mass according to these portions. Simply put, each cell looks at its neighborhood, computes the mass portions according to the affinities, and redistributes its own mass according to these portions.

Equation (\ref{eq:rhocomp}) simply encapsulates all the mass contributions to cell $(i,j)$. It will receive mass from all its neighbours, each time the portion being different, since each cell computes the portions according to its own neighborhood normalization. According to this explanation, it is easy to convince oneself this equation conserves mass; indeed, each cell simply redistributes its own mass to its neighbors each time step, and so mass is necessarily conserved. This can be shown by looking at the total mass $M^t = \sum_{i,j}\rho_{ij}^t$, as in (\ref{eq:massconservproof}) 
\begin{equations}
    M^{t+1} &= \sum_{i,j} \rho_{ij}^{t+1} = \sum_{i,j}\sum_{i',j'\in\mathcal{N}_{ij}}\frac{e^{\beta A_{ij}}}{Z_{i'j'}}\rho^t_{i'j'} \\ 
&= \sum_{i',j'}\frac{\rho^t_{i',j'}}{Z_{i'j'}}\sum_{i,j\in\mathcal{N}_{i'j'}}e^{\beta A_{ij}} = \sum_{i',j'}\frac{\rho^t_{i',j'}}{Z_{i'j'}}Z_{i'j'} = M^t\label{eq:massconservproof}
\end{equations}
To permute the sums, we used the fact that the neighborhood is 'reflexive', namely that $(i,j) \in \mathcal{N}(k,l) \Leftrightarrow (k,l) \in \mathcal{N}(i,j)$, which can be used to show the sums in (\ref{eq:massconservproof}) can be interchanged\footnote{$\sum_{i,j}\sum_{k,l\in \mathcal{N}(i,j)}=\sum_{i,j}\sum_{k,l}\mathds{1}_{k,l\in \mathcal{N}(i,j)}=\sum_{i,j}\sum_{k,l} \mathds{1}
_{i,j\in \mathcal{N}(k,l)}=\sum_{k,l}\sum_{i,j\in \mathcal{N}(k,l)}$}.

This evolution rule is captured in the broad family of mass conserving rules (\ref{eq:gen_massupdate}), where $I(x',x)$ is the mass portion going from $x'$ to $x$.
\begin{equation}
    \rho^{t+1}(x) = \sum_{x'} I(x',x)\rho^t(x'),\,\,\,\sum_{x} I(x',x)=1\label{eq:gen_massupdate}
\end{equation}
Where in our case, $I(x',x) = \frac{e^{\beta A(x)}}{Z(x')}\mathds{1}_{x'\in \mathcal{N}_x}$. Notably, Flow-Lenia (\cite{plantecFlowLeniaOpenendedEvolution2023a}) is another example of an update rule of the form (\ref{eq:gen_massupdate}), see Sec.\nameref{sec:flowcompare}.

\subsection{Continuous Limit}
One of the strengths of the update rule (\ref{eq:rhocomp}) is that it admits a continuous limit, both in time and space, as long as the affinity $A_{ij}$ is a smooth function of space in this limit. To display this, we must rewrite (\ref{eq:rhocomp}) to display the presence of the discretization parameters $\Delta t$ and $\Delta x$. To do so, we first re-arrange (\ref{eq:rhocomp}) into (\ref{eq:arrangerhocomp})
\begin{equation}
    \rho_{ij}^{t+\Delta t}-\rho_{ij}^t = \sum_{i',j'\in \mathcal{N}_{ij}} \frac{e^{\beta A_{ij}}}{Z_{i'j'}}\rho^t_{i'j'} - \rho^t_{i'j'} \label{eq:arrangerhocomp}
\end{equation}
Where we simply substracted $\rho_{ij}^t$ on both sides, and re-label $t+1\rightarrow t+\Delta t$ in order to make explicit the discretized time-derivative on the LHS. Now, we Taylor-expand the LHS in $\Delta t$, and the RHS in $\Delta x$ to leading order. After re-arranging, we obtain (see App. \ref{app:continuous_limit} for more steps) :
\begin{equation}
    \Delta t \partial_t\rho + O(\Delta t^2) = \frac{1}{3}\Delta x^2\left(\Delta \rho - 2\beta\nabla\cdot(\rho \nabla A) \right)+ O(\Delta x^4) \label{eq:continuousunnormalized}
\end{equation}
Equation (\ref{eq:continuousunnormalized}) tells us that the update rule will admit a continuous limit, so long as we correctly divide by $\Delta t$ and $\Delta x^2$ in the discretized equation. Consequently, the correctly rescaled discretization and its continuous limit are given in equation (\ref{eq:true_eqs}):
\begin{equations}
    \rho^{t+\Delta t}_{ij}&=\rho_{ij}^t(1-\frac{3\Delta t}{\Delta x^2})+\frac{3\Delta t}{\Delta x^2}\sum_{i',j'\in \mathcal{N}(i,j)} \frac{e^{\beta A_{ij}}}{Z_{i'j'}}\rho^t_{i'j'}\label{eq:true_eqs}\\
    \partial_t \rho &= \Delta\rho - 2 \beta \nabla \cdot (\rho \nabla A)
\end{equations}
We can see that in the original formula (\ref{eq:rhocomp}), we had (implicitly) chosen $3\Delta t = \Delta x^2$, with $\Delta x$ given by the grid size. Notably, $\frac{3\Delta t}{\Delta x^2}=1$ is a critical value, as the discretization breaks down above this point due to the negative pre-factor to $\rho_{ij}^t$. This limiting value is to be expected as even for pure diffusion with a similar full stencil, the stability condition requires $2\Delta t<\Delta x^2$. In practice, we always use $3\Delta t = \Delta x^2$, which both simplifies the evolution equation and maximizes the 'speed' of the dynamics by choosing the biggest possible $\Delta t$, given $\Delta x$.

\begin{figure}[h]
    \centering
    \includegraphics[width=0.7\linewidth]{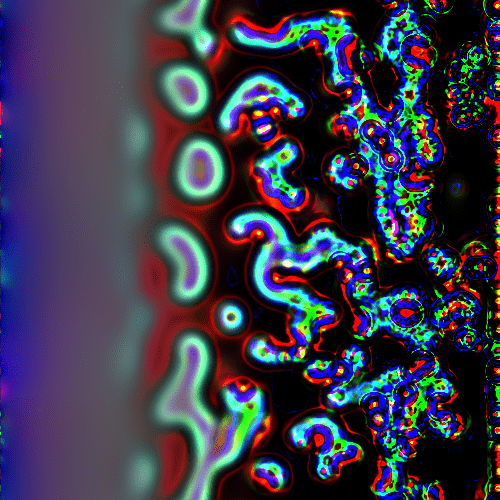}
    \caption{\footnotesize Display of the effect of $\beta$, growing from 0 to 10 going from left to right. At low $\beta$, diffusion dominates, while increasing $\beta$ generates sharper dynamics and more solitons.}
    \label{fig:betaindifflenia}
\end{figure}

Equation \ref{eq:true_eqs} de-mystifies the mass-conserving evolution rule, and the role of $\beta$. The rule effectively conserves mass by applying a combination of a fixed diffusion, and advection along the gradient of $A$. The $\beta$ parameter sets the relative strength of the two effects; at $\beta=0$ we have pure diffusion, and the effect of the advection becomes more prominent as we increase it (see Fig. \ref{fig:betaindifflenia} for display of the effect of $\beta$).

While $\beta$ can in some ways be interpreted as an inverse temperature (e.g. in the normalization in (\ref{eq:rhocomp}) it acts as the 'temperature' of the softmax, and as we increase it, the diffusion effect increases), this is simply a useful analogy rather than a rigorous identification. In particular, we note that we can take $\beta<0$ values without any problems, as this is simply equivalent to inverting the affinity flow direction.

One interesting limit to consider is $\beta\rightarrow \infty$, which is well-defined only in the discretized version of the evolution equations\footnote{Taking $\beta\rightarrow \infty$ in (\ref{eq:true_eqs}) breaks down as it implies infinite derivatives}. In this limit, each cell will give all of its mass to the biggest affinity neighbor. The evolution equation reduces to (\ref{eq:beta_infinite}).
\begin{equation}
    \rho^{t+1}_{ij}=\sum_{k,l\in \mathcal{N}_{ij}} \mathds{1}\left(A_{ij}=\max_{\mathcal{N}_{kl}}(A_{kl})\right)\rho^t_{kl} \label{eq:beta_infinite}
\end{equation}

This limit is very useful in the case where we want to use MaCE with an automaton with a discrete set of possible values, which precludes using (\ref{eq:rhocomp}) since it necessitates a continuous division of mass. Eq.(\ref{eq:beta_infinite}), however, only requires that the states of the automaton can be added together. In \nameref{sec:massivelife}, we use this rule to create a mass-conserving totalistic automaton (\cite{weissteinTotalisticCellularAutomaton}) with a Moore neighborhood.

\subsection{MaCELenia}
While we have shown that MaCE is a very versatile evolution rule that can be attached to most existing CA models, in this work we will focus our exploration chiefly on its fusion with Lenia, which we will call MaCELenia. As Lenia has already been extensively studied and shown to be able to support a wide variety of behaviours, we believe it is a good example to test our evolution rule. 

As a reminder, in Lenia, a 'growth' value $G_c(x,\rho)$ is computed for each pixel and each channel. First the matter fields $\rho_c$ are convolved with a set of kernels $K_{\tilde{c}c}$, which determine how channel $\tilde{c}$ influences $c$. Then the resulting values are fed through a non-linear functions $g_{\tilde{c}c}$ (the 'growth function'), before taking a weighted average with weights $W_{\tilde{c}}$. Equation (\ref{eq:growthlenia}) describes how the growth value is computed for each pixel.
\begin{equation}
    G_c = \sum_{\tilde{c}} W_{\tilde{c}} \,\,g_{\tilde{c}c}\left(K_{\tilde{c}c}*\rho_{\tilde{c}}\right)\label{eq:growthlenia}
\end{equation}
Usually, this growth value $G_c$ is then used to directly update the state $\rho_c$, adding or substracting matter from each channel according to the sign. In our case, we simply interpret this growth value as the affinity ($G_c=A_c$, in the notation of \ref{eq:true_eqs}), and apply the MaCE rule!

\begin{figure}
    \centering
    \includegraphics[width=0.8\linewidth]{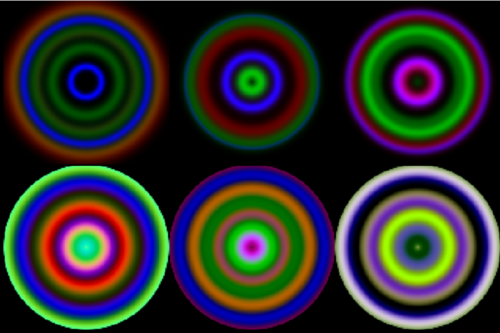}
    \caption{\footnotesize Example random kernels generated with different methods (top:Gaussian, Bottom:Fourier). Fourier series kernels are more flexible, but the parameter space is harder to navigate.}
    \label{fig:kernels}
\end{figure}

We will be using a slightly modified version of MultiChannel Lenia (\cite{chanLeniaExpandedUniverse2020a}), with altered methods of computing the Growth function and Kernels. The first generation method is close to the 'classic' Lenia. Each Kernel is rotationally symmetric, and its shape is determined by three sets of three real numbers $(\beta_k, \mu_k, \sigma_k)$ which define the shape and location of three Gaussian bumps $\beta_k \exp{-\frac{(x-\mu_k)^2}{2\sigma_k^2}}$. One such kernel is generated for every pair of channel interactions, resulting in $c^2$ Kernels for an automaton with $c$ channels. See fig. \ref{fig:kernels} for an example of such generated kernels. To sample random kernels, we simply randomize the three parameters using suitable ranges.

For the Growth function, we use the same unimodal Gaussian bump that is used in the original Lenia that depends only on two parameters, $\mu$ and $\sigma$, and is given by $2\exp(-\frac{(x-\mu)^2}{2\sigma^2})-1$\footnote{Note that in MaCELenia, we could drop the $-1$, as what matters is only differences of affinity but since it does not affect performance, we keep as it keeps closer to the original model}. For random search, we use both independent sampling of $\mu$ and $\sigma$, and a dependent sampling which sample $\sigma$ as to place the edge of the growth bump close to 0, which gave good results in Lenia (see (\cite{papadopoulos2024lookingcomplexityphaseboundaries})) for details).

To expand the parameter search space, we also experimented with generating more general (smooth) functions simply by sampling a finite number of Fourier coefficients and converting the result back to real space. We add some post-processing as necessary to rescale the function to the desired range, as well as clip values if needed (see the code repo for more details).

\subsection{Similarities with Flow-Lenia}\label{sec:flowcompare}
The update equation in (\ref{eq:rhocomp}) is reminiscent of the evolution rule used in Flow-Lenia (\cite{plantecFlowLeniaOpenendedEvolution2023a}). In Flow-lenia, a flow field is computed using the gradient of the affinity, then, matter is moved along this flow field using 'reintegration tracking' (\cite{morozReintegrationTracking}). The update rule can also be written as in (\ref{eq:gen_massupdate}). In Flow-Lenia, the effective 'neighborhood' (i.e. which $x'$s we consider) is technically the full domain, as mass could in principle flow in from abritrary distances. Still, a big similarity is that in both approaches, the main driving force is the advection along the gradient of the affinity, although the update rule itself is different. However, we think that MaCELenia provides several improvements.

The first advantage of MaCE is that it has a well-defined continuous limit, as we have shown. In contrast, reintegration tracking (\cite{morozReintegrationTracking}) is a complicated schema, that introduces additional discretization hyperparameters (such as $s$). It is not clear what the correct scaling of parameters is to obtain a well-defined continuous limit, if it exists.

The main advantage we believe lies in the simplicity and robustness of the update rule. Indeed, the rule can be stated as a straightforward 'cellular automaton-like' update, and there are no conditions on $A_{ij}$ for the mass conservation to hold. In Flow Lenia, the reintegration tracking forces conditions on the maximal possible values of the flow (should be on the order of the grid size), which will cause mass loss if they are too high (or flow will be clipped to prevent it, which will alter the dynamics).

In terms of results using MaCELenia, we find that dynamics in our schema are faster (at equivalent framerate) and produce solitons more easily, even though for the same Lenia parameters, the dynamics look qualitatively similar (which is expected due to the main driving force being the same in both cases). See App. \ref{app:compare_flow} for more details and direct comparisons of dynamics.

\subsection{Cross Channel Mass Extension}\label{sec:xchanmass}
Under our scheme, mass is conserved within channels, and mass cannot move from one channel to another. While this does not hinder a variety of interesting cross-channel interactions so long that the affinity is affected by all channels, for some applications, it can be desirable to allow cross-channel mass movement, see e.g., Sec.\nameref{sec:macenca}. The most natural way to do this is to simply extend the neighborhood of (\ref{eq:rhocomp}) to include the cells in all channels, and allow cells to redistribute their mass inside this bigger neighborhood. 

This cross-channel extension works well for MaCE-NCA, but in models like MaCELenia, the affinity is computed differently for each channel, producing highly discontinuous values across channels. In contrast, the affinity varies quite smoothly within a channel, due to being computed using smooth kernels. This has the undesirable effect that most of the mass is moved to the 'dominant' channel at each time step, and the dynamics become either flickery or concentrate all mass inside one channel. To remedy this, we can decouple the 'spatial' mass redistribution from the 'cross-channel' one. This allows us to tune the speed of the cross-channel sharing while remaining very close in spirit to \ref{eq:rhocomp}.

In this approach, we first begin by applying the spatial MaCE, unchanged from \ref{eq:rhocomp}, denoting its output as $\hat{\rho}_{ij}^{c,t+1}$. Then, for cross-channel sharing, we compute the affinity normalization $\hat{Z}_{ij}$ across channels using the same $\beta$, considering all channels to be in the same 'neighborhood', eq.(\ref{eq:affinitynormal}).
\begin{equation}
    \hat{Z}_{ij} =\sum_c e^{\beta A^c_{ij}} \label{eq:affinitynormal}
\end{equation}
Notice that the normalization $\hat{Z}_{ij}$ does NOT depend on the channel, because all channels have the same 'channel-neighborhood'. All that is left to do is apply the mass redistribution as in (\ref{eq:true_eqs}), replacing $\frac{3\Delta t}{\Delta x^2}\rightarrow \alpha$, which will become a tunable speed parameter, obtaining (\ref{eq:cross-chan-update}).
\begin{equation}
    \rho^{c,t+1}_{ij} = \hat{\rho}^{c,t+1}_{ij}(1-\alpha) + \alpha \frac{e^{\beta A^c_{ij}}}{\hat{Z}_{ij}}\sum_{\tilde{c}}\hat{\rho}^{\tilde{c},t+1}_{ij}
    \label{eq:cross-chan-update}
\end{equation}
Note that compared to (\ref{eq:true_eqs}), $\hat{Z}_{ij}$ can be taken out of the sum, because all channels have the same neighborhood. This allows to rephrase this equation in simpler terms: $\frac{e^{\beta A^c_{ij}}}{\hat{Z}_{ij}}$ are actually the 'target' mass portions that each channel wants to get at. Hence, the target mass for each channel is this portion multiplied by the total mass currently available in all channels. Eq. (\ref{eq:cross-chan-update}) then displaces mass at rate $\alpha$ towards the target. For MaCELenia, good values are $0.01<\alpha<0.1$.

\section{MaCELenia  Capabilities}
In this section, we demonstrate the capabilities of our mass-conserving scheme with regard to its interactions with Lenia. 


We highlight four important capabilities of MaCELenia: the ease with which we find solitons,  the diversity found in one parameter set, intrinsic competition in resource-constrained environments, and the ability to optimize parameters towards a goal. 

\subsection{Results from Random Sampling}
To first gauge the capabilities of MaCELenia, we simply observe what sort of dynamics can be obtained by randomly sampling parameters and initial conditions. We are immediately surprised by the variety of different dynamics, and the ubiquity of interesting behaviours just from random sampling. In this section, we try to quantify this feeling by counting how often solitons appear in the dynamics, although a better way to get a feeling for them is to watch some videos at \href{https://etimush.github.io/MaCE-Videos/}{the video page}.

We go through one hundred random generations of MaCELenia. Each generation starts with a random circle of Perlin noise in the center, and we let it evolve for 1000 iterations, after which we visually determine whether the system produces solitons. We define solitons qualitatively as moving 'creatures', namely some 'pockets' of mass that can move around.

\begin{figure}[h]
\centering
    \begin{subfigure}[b]{0.48\linewidth}
        \centering
         \includegraphics[width=\linewidth]{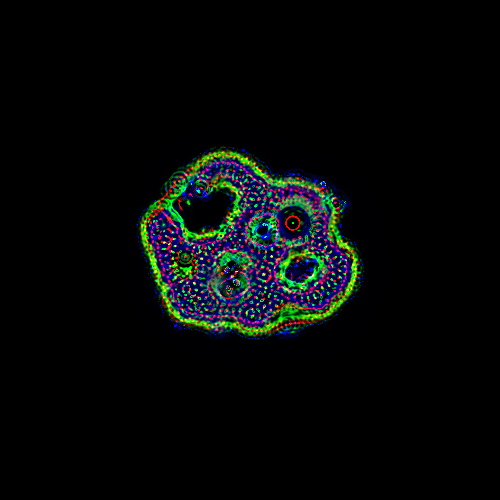}
         \caption{A parameter set considered to not produce solitons}
         \label{noSoli}
    \end{subfigure}
    \begin{subfigure}[b]{0.48\linewidth}
        \centering
        \includegraphics[width=\linewidth]{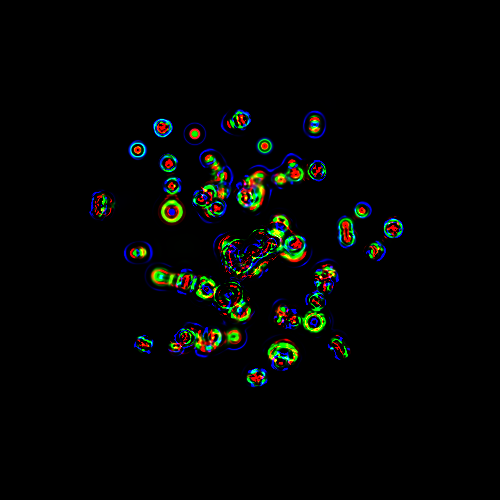}
         \caption{A parameter set considered to produce solitons}
         \label{soli}
    \end{subfigure}
\caption{\footnotesize Soliton selection criteria.}
\label{fig:soliSelecCrit}
\end{figure}

Figure \ref{fig:soliSelecCrit} displays two examples of systems with and without solitons. On the left (Figure \ref{noSoli}), we see a connected mass with little differentiation between sections. On the right (Figure \ref{soli}), we see a plethora of solitons forming, with clear separation between them. After 100 random generations, we found that $\sim 83\%$ of the time the system has soliton emergence.

Note that the sampling method we used does not sample all parameters independently; we use a sampling method found in (\cite{papadopoulos2024lookingcomplexityphaseboundaries}), which produced better results in plain Lenia (see App.\ref{app:random_sampling}). We tried independent sampling as well, which produced less solitons ($\sim 61\%$) but with more variety and color mixing. We also tried with kernels generated by sampling Fourier coefficients (see Fig.\ref{fig:kernels}, as well as App.\ref{app:randomfourier} for details), which produces qualitatively similar results but has the potential for more variety.

\vspace{-.5cm}
\subsection{Diversity in One Parameter Set}
One unexpected property of MaCELenia is the diversity in morphologies that can be found in a single parameter set. This is in stark contrast to Lenia, where most often a single parameter set supports the existence of only one (rarely two/three) kind of solitons. In MaCELenia, it is not uncommon to have dozens of different kinds of moving solitons in a single parameter set. 

\begin{figure}[h]
\centering
    \begin{subfigure}[b]{0.48\linewidth}
        \centering
         \includegraphics[width=\linewidth]{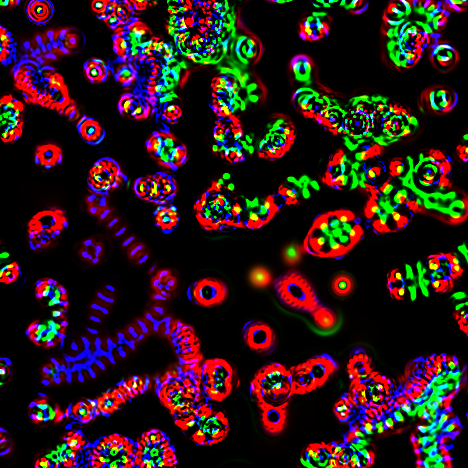}
    \end{subfigure}
    \begin{subfigure}[b]{0.48\linewidth}
        \centering
         \includegraphics[width=\linewidth]{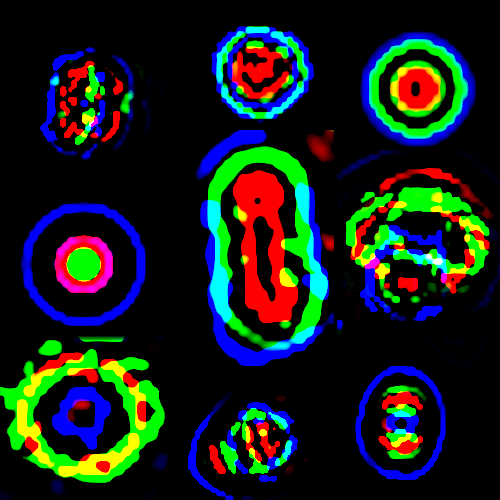}
    \end{subfigure}
\caption{\footnotesize \textbf{Left}: snapshot of MaceLenia world after 100 steps. \textbf{Right}: selection of solitons found in one parameter set.}
\label{fig:DivOneSt}
\end{figure}

Figure \ref{fig:DivOneSt}, on the left, depicts a world after 100 evolution steps, all started on random low-frequency Perlin noise. On visual inspection, we see a wide variety of patterns emerge, including solitons and "blobs" of mass. On the right, we show a selection of nine solitons found in a single set of parameters, all originating from the same Perlin initialization. These solitons were stable over time and only fluctuated when in contact with other solitons.

\subsection{Intrinsic Evolution}
As we have seen, a plethora of solitons can emerge inside a MaCELenia world. This variety is a necessary condition for evolution, so in this section, we probe whether there can be some hints of intrinsic evolution in MaCELenia.

Although we could simply let a simulation run and see whether some solitons are selected, it is more interesting to exploit the finite resource which is mass to introduce competition for it. To do so, we initialize a random MaCELenia world, and introduce a constant mass decay, namely all channels loose mass at a constant rate (similar to \cite{plantecFlowLeniaOpenendedEvolution2023a})). To avoid ending up with an empty world, the removed mass is periodically reintroduced as 'food pellets', in a separate 'food channel', in such a way that the total mass (food + CA channels) remains constant. The way MaCELenia creatures can sense and consume the food is simple; CA channels sense the food channel through the same convolution kernels, and the food mass is transferred to the CA channels whenever they overlap, as long as the CA mass exceeds a certain threshold \footnote{We introduced the minimum mass threshold to consume food to prevent mass being consumed by very low concentrations, which are common because of the diffusion in the MaCE. However, this sometimes still happens, as the food sensing can cause low-density mass to concentrate.}

\begin{figure}[h]
\centering
    \begin{subfigure}[b]{0.32\linewidth}
        \centering
         \includegraphics[width=\linewidth]{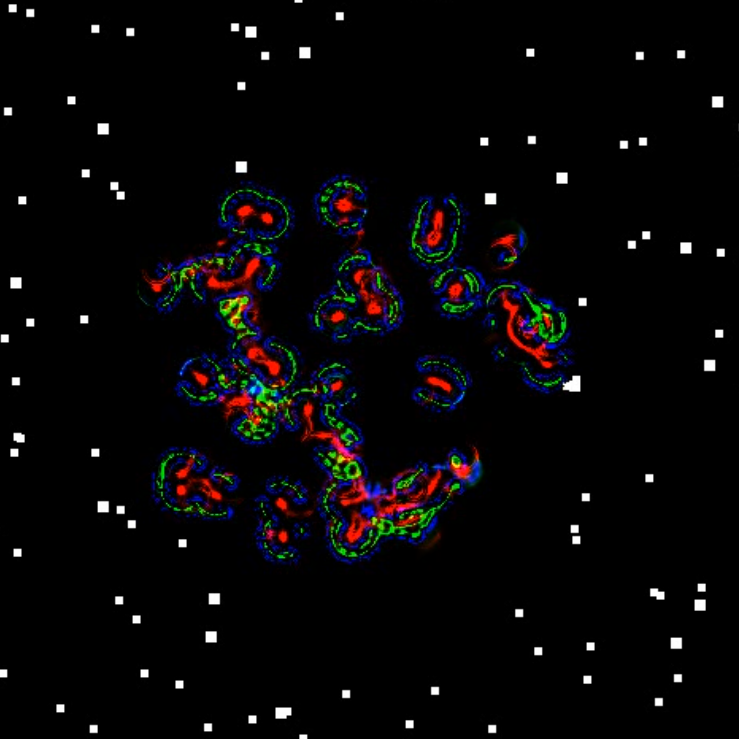}
    \end{subfigure}
    \begin{subfigure}[b]{0.32\linewidth}
        \centering
         \includegraphics[width=\linewidth]{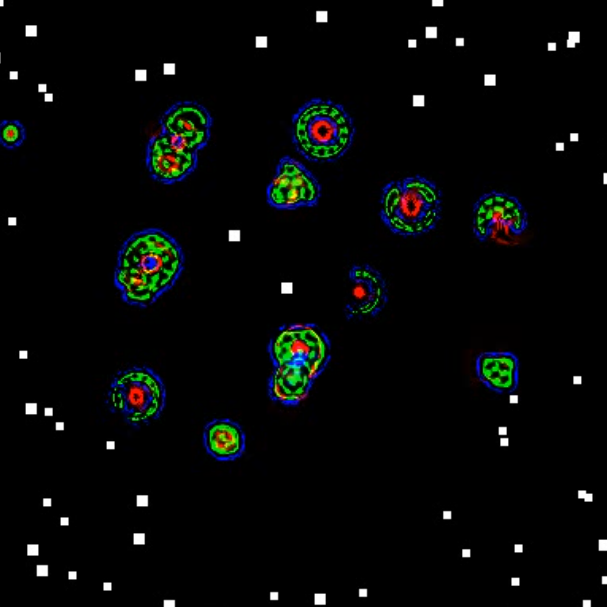}
    \end{subfigure}
    \begin{subfigure}[b]{0.32\linewidth}
        \centering
         \includegraphics[width=\linewidth]{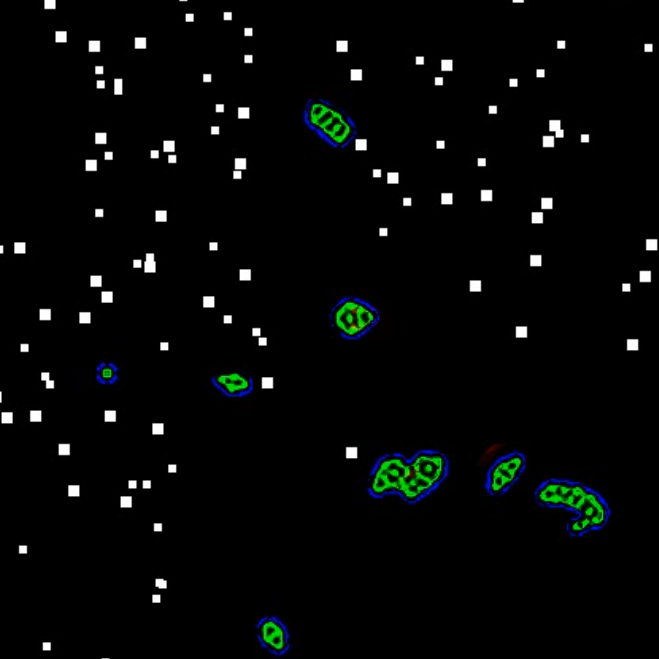}
    \end{subfigure}
\caption{\footnotesize Intrinsic evolution experiment, time increases to the right}
\label{fig:IntrinsicComp}
\vspace{-0.2cm}
\end{figure}

Figure \ref{fig:IntrinsicComp} shows the time evolution of a MaCELenia system in a mass-decaying world simulated for $\sim10$k steps, in white are the food pellets. Shortly after initialization, we observe a semi-static species consisting of a red nucleus. Failing to move, mass decays rapidly, and the configurations of the soliton change, expelling the red nucleus. This triggers a transition that makes the solitons highly mobile, and they start to colonize the space and reproduce as they gather extra mass. When mass is consumed, red is reintroduced, which sometimes freezes the solitons, which begin foraging anew when mass decays again.


\begin{figure}[h]
\centering
    \begin{subfigure}[b]{0.48\linewidth}
        \centering
         \includegraphics[width=\linewidth]{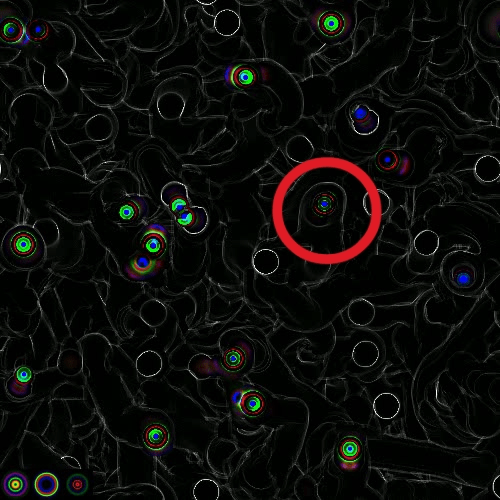}
         \caption{Stage 1}
         \label{stage 1}
    \end{subfigure}
    \begin{subfigure}[b]{0.48\linewidth}
        \centering
         \includegraphics[width=\linewidth]{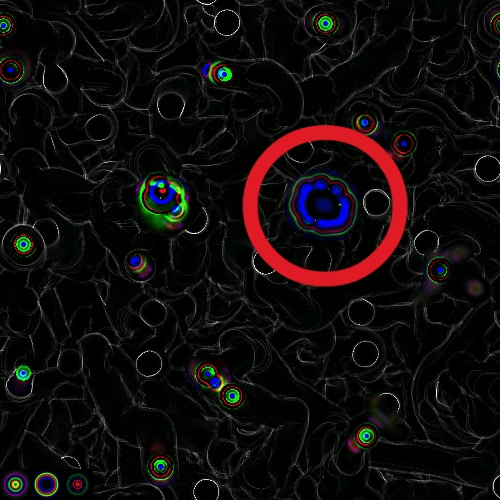}
         \caption{Stage 2}
         \label{stage 2}
    \end{subfigure}
    \\
    \begin{subfigure}[b]{0.48\linewidth}
        \centering
         \includegraphics[width=\linewidth]{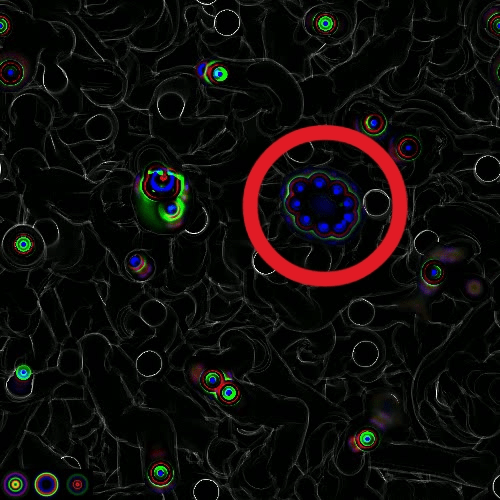}
         \caption{Stage 3}
         \label{stage 3}
    \end{subfigure}
    \begin{subfigure}[b]{0.48\linewidth}
        \centering
         \includegraphics[width=\linewidth]{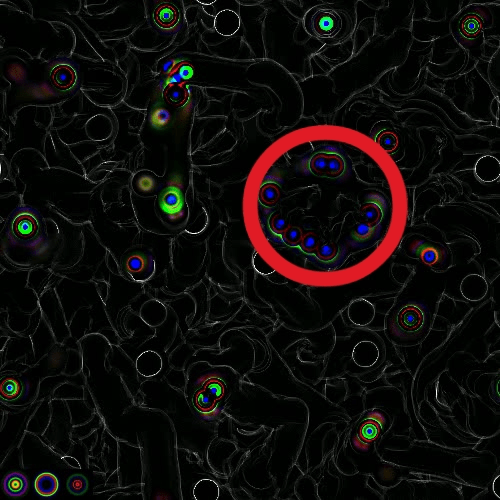}
         \caption{Stage 4}
         \label{stage 4}
    \end{subfigure} 
\caption{Mitosis}
\label{fig:Mitosis}
\vspace{-0.3cm}
\end{figure}

Figure \ref{fig:Mitosis} depicts an instance of a dominant soliton that undergoes "mitosis". Here, we performed a slightly different experiment; mass still decays over time, but is instantly added back as food where it decayed (shown in the white shapes). In stage 1 (Figure \ref{stage 1}) the soliton undergoes a phase change into the mitotic state. Stage 2 (Figure \ref{stage 2}) follows as the mass of the soliton expands, followed by stage 3 (Figure \ref{stage 3}) where the mass differentiates back into many solitons. Finally, stage 4 (Figure \ref{stage 4}), where the solitons go their own ways to repeat the cycle. 

These experiments confirm that the variety of creatures that can emerge in MaCELenia can support some primitive intrinsic selection. The main missing piece towards evolution is the capacity to transmit their 'genes': in these experiments, each new creature goes through the selection process each time! Bigger and longer experiments are needed to see how far this can be pushed!
\subsection{Extrinsic Evolution}
In this section, we detail two small experiments on extrinsically evolving MaCELenia parameters towards a set goal. In the first example, we set up a world with harsh mass decay. As before, mass that has decayed is periodically added back into the world at random locations in the form of food. The MaCELenia system tries to maximize its total mass by the 1000th step. To do this, creatures must find and eat food as fast as possible to counteract the mass decay. In the second example, a MaCELenia creature evolves to move from the center of the world to the bottom right corner. 

We use Evolutionary Strategies (ES) \cite{Beyer2002} to mutate the kernel and growth parameters. The population of individuals (different parameter sets) is 16.

\begin{figure}[h]
\centering
    \begin{subfigure}[b]{0.32\linewidth}
        \centering
         \includegraphics[width=\linewidth]{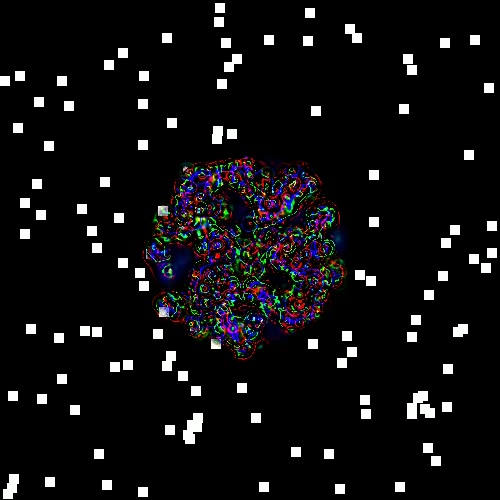}
         \caption{t = 30}
         \label{step30evo}
    \end{subfigure}
    \begin{subfigure}[b]{0.32\linewidth}
        \centering
         \includegraphics[width=\linewidth]{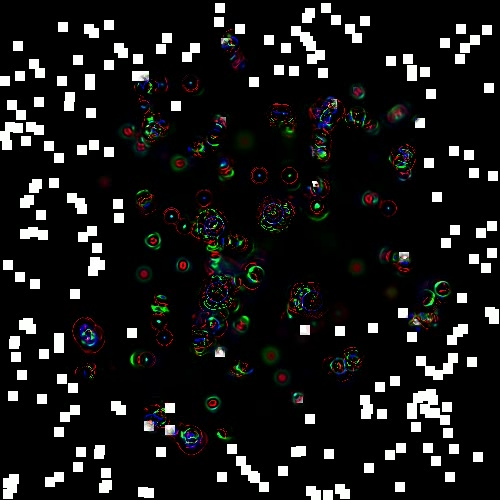}
         \caption{t = 300}
         \label{sep300evo}
    \end{subfigure}
    \begin{subfigure}[b]{0.32\linewidth}
        \centering
         \includegraphics[width=\linewidth]{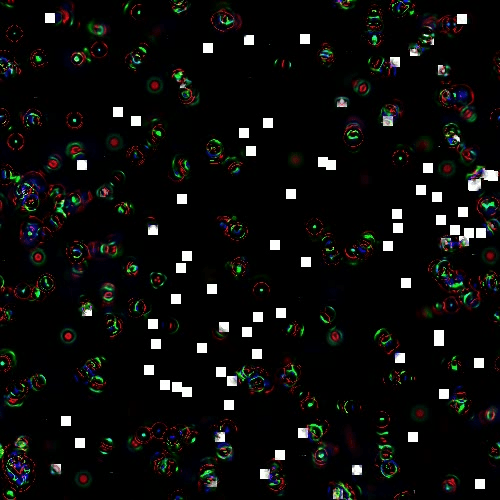}
         \caption{t = 1000}
         \label{step999evo}
    \end{subfigure}
\caption{\footnotesize Time evolution of the final fittest individual for a system evolved to maximize mass in a harsh environment.}
\label{fig:EvoStages}
\end{figure}

Figure \ref{fig:EvoStages} depicts the time evolution of the final fittest individual of the evolutionary process. Since mass decay is extremely harsh, to survive the system must produce fast-moving solitons that replicate easily. At $t = 30$ (Figure \ref{step30evo}), the system starts its transformation from noise to solitons. At $t = 300$ (Figure \ref{sep300evo}), a variety of solitons become apparent. By $t = 1000$ (Figure \ref{step999evo}), the solitons have spread out completely, attempting to maximize the mass they consume. 

\begin{figure}[h]
\centering
    \begin{subfigure}[b]{0.32\linewidth}
        \centering
         \includegraphics[width=\linewidth]{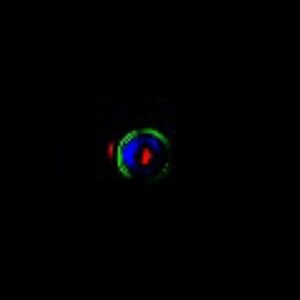}
         \caption{t = 10}
         \label{step30move}
    \end{subfigure}
    \begin{subfigure}[b]{0.32\linewidth}
        \centering
         \includegraphics[width=\linewidth]{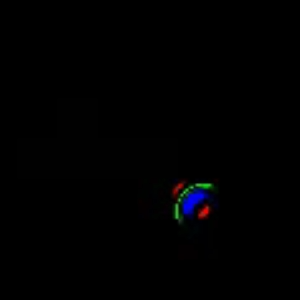}
         \caption{t = 200}
         \label{sep200move}
    \end{subfigure}
    \begin{subfigure}[b]{0.32\linewidth}
        \centering
         \includegraphics[width=\linewidth]{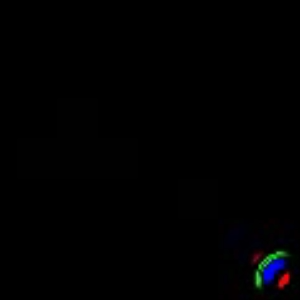}
         \caption{t = 500}
         \label{step499move}
    \end{subfigure}
\caption{\footnotesize Time evolution of the final fittest individual for a system evolved to move to the bottom right corner.}
\label{fig:EvoStagesMove}
\vspace{-0.3cm}
\end{figure}

Figure \ref{fig:EvoStagesMove} shows a MaCELenia creature that was evolved to move from the center of the world to the bottom right in 500 time steps. In both experiments, finding high fitness solutions was painless and required little to no hyperparameter tuning. This shows MaCELenia has a rich and dense array of behaviours in its parameter set, that can be easily discovered through genetic algorithms.

\section{Versatility of MaCE}
In this section, we present the ease of portability of MaCE, which can be attached to most existing CAs with great results. To demonstrate this near "plug-and-play" nature, we transform two Cellular Automata to be mass-conserving - a Neural Cellular Automaton and a discrete CA, and detail the minimal changes necessary for them to work. 
\subsection{MaCE-NCA}\label{sec:macenca}
NCAs are a class of CA characterized by Neural-network-based cell-state update functions, which are not mass-conserving. In (\cite{mordvintsevGrowingNeuralCellular2020}), the growth mechanism was mediated by a special $\alpha$ channel, where cells were considered "dead" if either their own $\alpha$ channel or their neighbors' $\alpha$ channels failed to meet a threshold.

With mass conservation, this mechanism is no longer needed, as previously dead (empty) cells need to acquire mass from their immediate neighbors. For NCA, the most suitable use of MaCE is in the $0 < \beta <1$ range, with mass conservation only applied to the visible (RGB) channels. The low $\beta$ allows the NCA to finely choose how fast to redistribute the mass by tuning the affinity differences.

To transform an NCA into a MaCE-NCA, we perform the same steps as (\cite{mordvintsevGrowingNeuralCellular2020}) with one simple modification. All non-RGB channels are updated normally (without MaCE) by their respective update tensor computed by the Neural Network. The RGB channels are instead updated using MaCE, where the affinity is simply the RGB portion of the NCA update tensor.

We also need to take into account the color distribution of the final image. Indeed, it follows by conservation that the RGB channels of the seed cell must contain all the mass from which the final morphology grows. Thus, we are met with two options: allocate the correct mass proportion to each channel at the start, or allocate the mass equally between channels and allow for cross-channel redistribution. We choose to allow for cross-channel redistribution, doing so by extending the neighborhood of (\ref{eq:rhocomp}) as described in sec. \nameref{sec:xchanmass}

We use the AdamW optimizer, with $lr = 10^{-3}$ and a $66\%$ reduction in learning rate every 3000 backpropagation steps.

\begin{figure}[h]
\centering

    \begin{subfigure}[b]{0.32\linewidth}
        \centering
         \includegraphics[width=\linewidth]{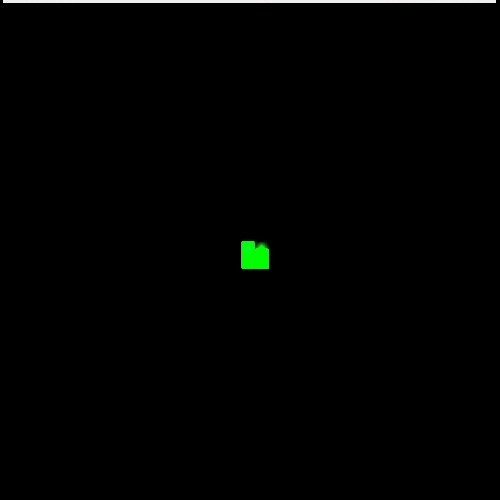}
         \caption{t = 5}
         \label{liz5}
    \end{subfigure}
    \begin{subfigure}[b]{0.32\linewidth}
        \centering
         \includegraphics[width=\linewidth]{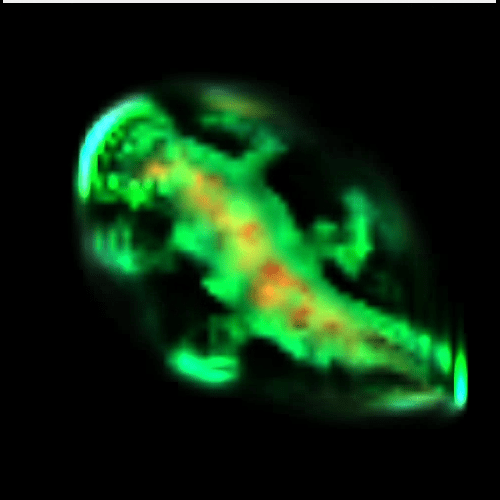}
         \caption{t = 32}
         \label{liz32}
    \end{subfigure}
    \begin{subfigure}[b]{0.32\linewidth}
        \centering
         \includegraphics[width=\linewidth]{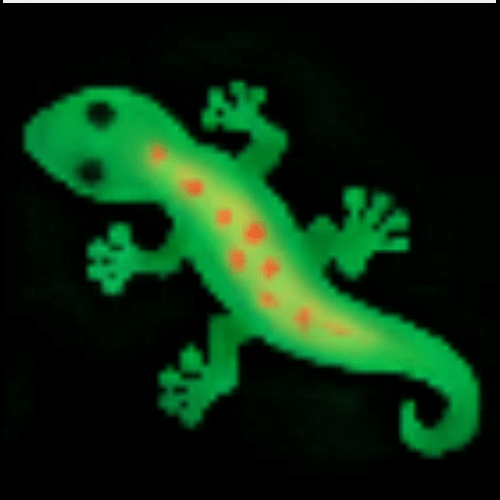}
         \caption{t = 96}
         \label{liz96}
    \end{subfigure}

\caption{Lizard Growing Mass Conservation}
\label{fig:growingliz}
\end{figure}

Figure \ref{fig:growingliz} shows the time evolution of a fully trained MaCE NCA growing from a seed-cell into a lizard. The CA starts off by transferring all its mass to the green channel (Figure \ref{liz5}); We presume this is because the color is the largest source of loss in the beggining. As the mass expands outwards (Figure \ref{liz32}), the shape of the lizard can start to be seen, alongside local differentiation of colors. By $t = 96$ (Figure \ref{liz96}), the lizard is fully formed with no mass to spare. 

We also performed experiments with the NCA variant IsoNCA\cite{mordvintsev_growing_2022} (see video page). We leave for future work the exploration of MaCENCA in more detail, in particular the continuous limit as in \cite{pajouheshgar_noisenca_2024}.
\subsection{MaCE Discrete CA} \label{sec:massivelife}
In this section, we provide an example of how MaCE can be applied even to discrete CAs. In a discrete CA, the most suitable use of MaCE is in the $\beta \rightarrow \infty$ limit as in (\ref{eq:beta_infinite}). This is because with discrete states, we cannot select arbitrary portions of mass to redistribute. Additionally, grainy dynamics are expected, so removing the 'diffusive' part of the rule seems reasonable.

One further constraint to apply (\ref{eq:beta_infinite}) is to be able to add states together. In this example, we do this in the simplest way: the states of our automaton are the natural numbers, and so we can interpret each cell state as being described by the number of 'mass quantas' it is carrying. A further choice must be made in applying (\ref{eq:beta_infinite}), and that is how to deal with degenerate maximal values of the affinity in the neighborhood (which are bound to happen because of the discrete state-space). In our implementation, we opt for a deterministic option, which is to divide equally the mass to be shared. Since the values are discrete, we perform an integer division, and the leftover mass simply remains in the central cell. Another possibility is to randomize among the maximal affinities where the full mass will go. This preserves the discrete spirit, but makes the dynamics stochastic.

With this in mind, the only thing left to describe is how we choose to compute the affinity. We work with the standard $3\times 3$ Moore neighborhood, with a holistic rule; each cell computes the total mass in the neighborhood, and feeds this to a non-linear growth function to obtain the affinity. 

\begin{figure}[h!]
\centering
    \begin{subfigure}[b]{0.46\linewidth}
        \centering
         \includegraphics[width=\linewidth]{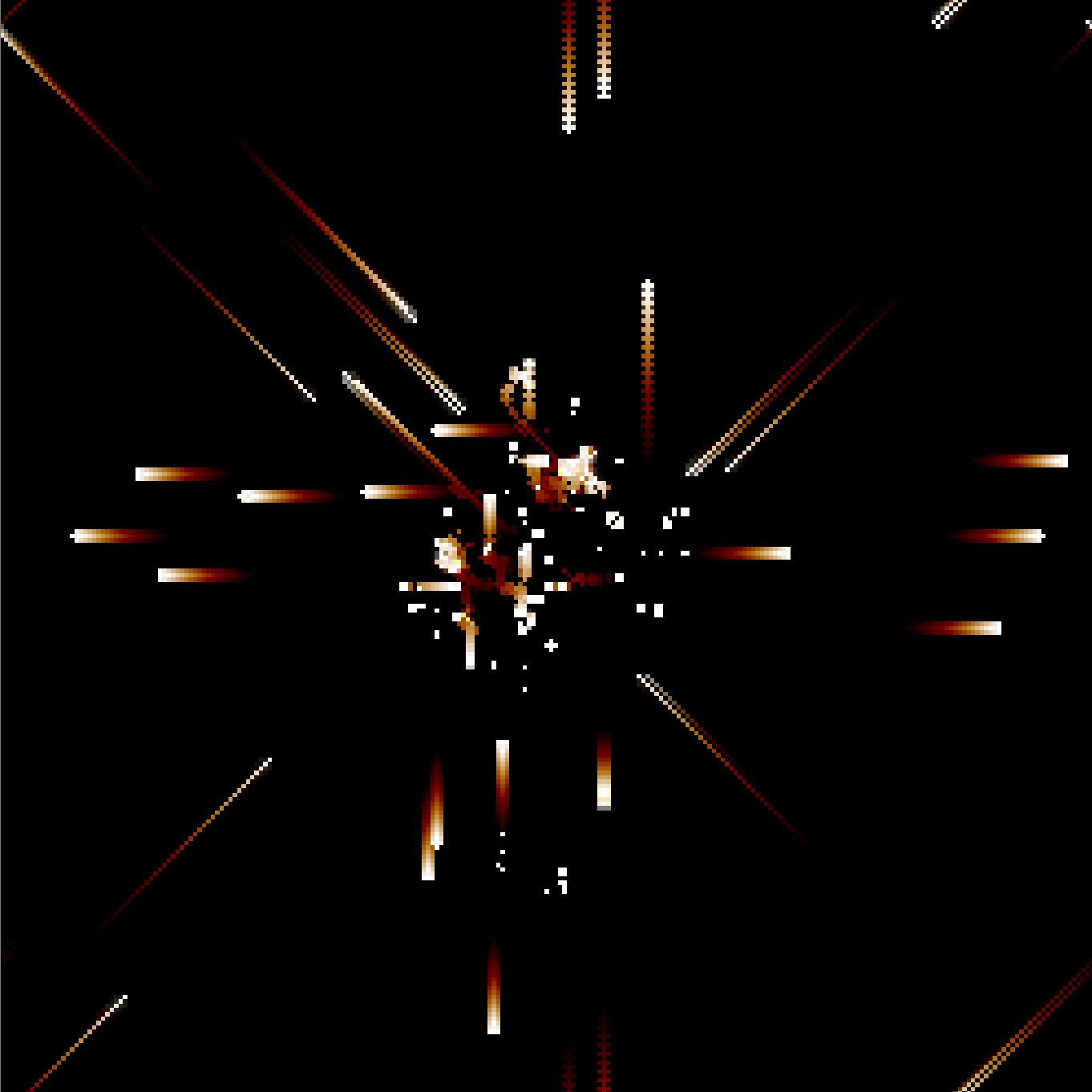}
    \end{subfigure}
    \begin{subfigure}[b]{0.46\linewidth}
        \centering
         \includegraphics[width=\linewidth]{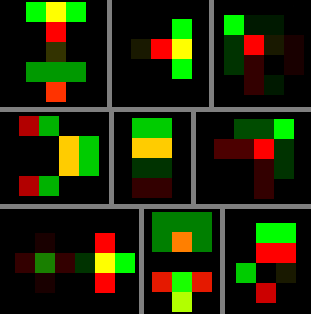}
    \end{subfigure}
    \caption{\footnotesize \textbf{Left}: display of glider diversity from random initialization circle. We display red trails that help distinguish the gliders and their speeds and directions. \textbf{Right}: Close view of gliders. We display the last and next-to-last states in green and red, respectively.}
    \label{fig:maceca}
\end{figure}

We test simple growth functions composed of a single gaussian bump. The qualitative impressions are similar to MaCELenia, in the sense that we very easily obtain a wide variety of gliders in a single ruleset. In Figure \ref{fig:maceca}, we display the configuration after initialization with a circle of noise, and the various 'gliders' that can be found in this world. In particular, we observe a plethora of non-trivial glider-glider interactions (bouncing, combining, destruction) which place the dynamics of this automaton firmly in class IV (\cite{wolframUniversalityComplexityCellular1984}), and we are confident that a Turing machine can be built using these dynamics.

Overall, from our modest exploration, the combination of MaCE with CA has high potential, and we leave for future work a more thorough search of what can be done by considering other ways of computing the affinity.

\section{Discussion}
In this paper, we introduced MaCE, an evolution rule that takes in real values (the affinities) as input and outputs a state update that conserves mass. MaCE is simple to implement, efficient, and is compatible with most existing CAs nearly as a 'slot-in' upgrade. To demonstrate this, we tested MaCE on Lenia, NCA, and discrete automata, and in each case, we have found it to yield very promising results. In particular, we analyzed it through MaCELenia in more detail and were very impressed by the consistency and variety of emergent solitons, especially within a single set of parameters.

There are some further extentions and modifications of MaCE that could be explored in the future. For instance, one shortcoming of MaCE is that the mass redistribution happens in a small $3\times 3$ neighborhood. Though this is fine for local rules, in cases like Lenia where the dynamics are actually non-local (the kernel has an extended size), they might look like they 'slow down' as we make the grid finer. This is perfectly consistent with the continuous limit (\ref{eq:true_eqs}) \footnote{When we lower $\Delta x$, $\Delta t$ must be decreased, and hence more timesteps are necessary to evolve the system by the same amount of time!}, but can make the dynamics too slow at small grid sizes ($\Leftrightarrow$ large kernel sizes). One could experiment with non-local version of MaCE, where mass is allowed to be distributed throughout larger neighborhoods. The continuous limit won't be expressible by a simple differential equation anymore, but it might yield surprising results!

Another interesting extension we would like to consider in the future would be some type of 'parameter sharing', or 'genome transport' where some parameters of the underlying CA (that are used to compute affinities), can also be transported and distributed around, similar to some experiments performed in Flow-Lenia. Such a scheme would allow the intrinsic evolution experiments to be pushed much further, as the diversity of emerging creatures would be that much higher. However, such a parameter sharing scheme would probably require a more thorough reworking of the underlying CA, which would be at odds with the ease of applicability of MaCE, and it is why we leave this exploration for future research!

Beyond that, we feel that we have only scratched the surface of what MaCE can produce when paired with existing systems. Taking the CA example alone, we have tested only holistic automata, with a fairly simple affinity function. The possibilities for the choices of affinities are literally endless, and the regularization from the mass conservation makes the density of interesting rules much higher. This makes us certain that a hidden gem is hiding somewhere in parameter space! In MaCELenia (but also in general) of particular interest would be to look for behaviour that display long range order and communication. Indeed, such behaviours are the necessary building blocks for complex self-organization at the next scale. With mass conservation this would involve the emergence of 'communication channels' where packets of mass would be exchanged for communication, which is reminiscent of the use of ion channels in neuronal interactions. To find such rules, the most promising path is using evolutionary algorithms with handcrafted rewards functions.

All in all, we hope that MaCE will allow the discovery of a plethora of new Artificial Worlds, bringing us yet another inch closer to Artificial Life.

\section{Acknowledgements}
We would like to thank Clément Hongler, Barbora Hudcová, Eugène Bergeron, Ehsan Pajouheshgar, Simon Ben Arous, Joao Penedones, Pauline Lamotte, Slackermanz, Bert Chan, José Boura de Matos for interesting discussions, suggestions and beta testing. This work was partially supported by an EPFL FSB Seed Funding Grant.
{\footnotesize
\bibliographystyle{apalike}
\bibliography{sample} 

\begin{thebibliography}{}

\bibitem[Beyer and Schwefel, 2002]{Beyer2002}
Beyer, H.-G. and Schwefel, H.-P. (2002).
\newblock Evolution strategies -- a comprehensive introduction.
\newblock {\em Natural Computing}, 1(1):3--52.

\bibitem[Chan, 2020a]{chanLeniaExpandedUniverse2020}
Chan, B. W.-C. (2020a).
\newblock Lenia and {{Expanded Universe}}.
\newblock {\em The 2020 Conference on Artificial Life}, pages 221--229.

\bibitem[Chan, 2020b]{chanLeniaExpandedUniverse2020a}
Chan, B. W.-C. (2020b).
\newblock Lenia and {{Expanded Universe}}.
\newblock {\em The 2020 Conference on Artificial Life}, pages 221--229.

\bibitem[Cook, 2004]{cookUniversalityElementaryCellular2004}
Cook, M. (2004).
\newblock Universality in {{Elementary Cellular Automata}}.
\newblock {\em Complex Systems}, page~40.

\bibitem[ereb0s labs, 2022]{ereb0slabs2022}
ereb0s labs (2022).
\newblock Tweet by ereb0s\_labs.
\newblock https://x.com/ereb0s\_labs/status/1601224524883972097.

\bibitem[Evans, 2001]{evansLargerLifeDigital2001}
Evans, K.~M. (2001).
\newblock Larger than {{Life}}: {{Digital Creatures}} in a {{Family}} of {{Two-Dimensional Cellular Automata}}.
\newblock {\em Discrete Mathematics \& Theoretical Computer Science}, DMTCS Proceedings vol. AA, Discrete Models: Combinatorics, Computation, and Geometry (DM-CCG 2001)(Proceedings).

\bibitem[Faldor and Cully, 2024]{faldor2024artificialopenendedevolutionlenia}
Faldor, M. and Cully, A. (2024).
\newblock Toward artificial open-ended evolution within lenia using quality-diversity.

\bibitem[Gardner, 1970]{MathematicalGamesFantastic1970}
Gardner (1970).
\newblock Mathematical {{Games}} - {{The}} fantastic combinations of {{John Conway}}'s new solitaire game "life".
\newblock page~6.

\bibitem[Heinemann, 2025]{heinemannChrxhAlien2025}
Heinemann, C. (2025).
\newblock Chrxh/alien.

\bibitem[Hickinbotham and Stepney, 2015]{hickinbothamConservationMatterIncreases2015}
Hickinbotham, S.~J. and Stepney, S. (2015).
\newblock Conservation of matter increases evolutionary activity.
\newblock In {\em European {{Conference}} on {{Artificial Life}} 2015}, pages 98--105. MIT Press.

\bibitem[Johnston and Greene, 2022]{johnston2022conways}
Johnston, N. and Greene, D. (2022).
\newblock {\em Conway's Game of Life: Mathematics and Construction}.
\newblock Self-published.
\newblock Hardcover, color printing, US letter (8.5 × 11 in).

\bibitem[Mohr, 2025]{mohrTommohrParticlelifeapp2025}
Mohr, T. (2025).
\newblock Tom-mohr/particle-life-app.

\bibitem[Mordvintsev et~al., 2022a]{mordvintsevParticleLeniaEnergybased}
Mordvintsev, A., Niklasson, E., and Randazzo, E. (2022a).
\newblock Particle {{Lenia}} and the energy-based formulation.
\newblock https://google-research.github.io/self-organising-systems/particle-lenia/.

\bibitem[Mordvintsev et~al., 2022b]{mordvintsev_growing_2022}
Mordvintsev, A., Randazzo, E., and Fouts, C. (2022b).
\newblock Growing {Isotropic} {Neural} {Cellular} {Automata}.
\newblock arXiv:2205.01681 [cs].

\bibitem[Mordvintsev et~al., 2020]{mordvintsevGrowingNeuralCellular2020}
Mordvintsev, A., Randazzo, E., Niklasson, E., and Levin, M. (2020).
\newblock Growing {{Neural Cellular Automata}}.
\newblock {\em Distill}, 5(2):e23.

\bibitem[Moroz, 2020]{morozReintegrationTracking}
Moroz, M. (2020).
\newblock Reintegration tracking.
\newblock https://michaelmoroz.github.io/Reintegration-Tracking/.

\bibitem[Pajouheshgar et~al., 2024]{pajouheshgar_noisenca_2024}
Pajouheshgar, E., Xu, Y., and Süsstrunk, S. (2024).
\newblock {NoiseNCA}: {Noisy} {Seed} {Improves} {Spatio}-{Temporal} {Continuity} of {Neural} {Cellular} {Automata}.
\newblock MIT Press.

\bibitem[Papadopoulos et~al., 2024]{papadopoulos2024lookingcomplexityphaseboundaries}
Papadopoulos, V., Doat, G., Renard, A., and Hongler, C. (2024).
\newblock Looking for complexity at phase boundaries in continuous cellular automata.

\bibitem[Plantec et~al., 2023]{plantecFlowLeniaOpenendedEvolution2023a}
Plantec, E., Hamon, G., Etcheverry, M., Oudeyer, P.-Y., {Moulin-Frier}, C., and Chan, B. W.-C. (2023).
\newblock Flow-{{Lenia}}: {{Towards}} open-ended evolution in cellular automata through mass conservation and parameter localization.

\bibitem[Reynolds, 1987]{reynoldsFlocksHerdsSchools1987}
Reynolds, C.~W. (1987).
\newblock Flocks, herds and schools: {{A}} distributed behavioral model.
\newblock {\em SIGGRAPH Comput. Graph.}, 21(4):25--34.

\bibitem[{Slackermanz}, 2023]{slackermanzMNCACompilationSlackermanz2023}
{Slackermanz} (2023).
\newblock {{MNCA Compilation}}.

\bibitem[Succi, 2001]{succi2001lattice}
Succi, S. (2001).
\newblock {\em The Lattice Boltzmann Equation, for Fluid Dynamics and Beyond}.
\newblock Oxford Science Publications.
\newblock Chapter 2 is about lattice gas Cellular Automata.

\bibitem[von Neumann and Burks, 1966]{vonneumann1966theory}
von Neumann, J. and Burks, A.~W. (1966).
\newblock {\em Theory of Self-Reproducing Automata}.
\newblock University of Illinois Press, Urbana.

\bibitem[Weisstein, 2002]{weissteinTotalisticCellularAutomaton}
Weisstein, E.~W. (2002).
\newblock Totalistic {{Cellular Automaton}}.
\newblock https://mathworld.wolfram.com/TotalisticCellularAutomaton.html.

\bibitem[Wolfram, 1984]{wolframUniversalityComplexityCellular1984}
Wolfram, S. (1984).
\newblock Universality and complexity in cellular automata.
\newblock {\em Physica D: Nonlinear Phenomena}, 10(1):1--35.

\end{thebibliography}
}
\clearpage
\appendix
\onecolumn

\section{Derivation for the continuous version of MaCE}\label{app:continuous_limit}
In this appendix, we go through more thorough steps to recover the continuous limit of MaCE, as in (\ref{eq:true_eqs}). We start from eq.(\ref{eq:arrangerhocomp}). The first step, is Taylor-expanding the LHS, to explicit the time derivative :
\begin{equation}
    \Delta t \frac{\partial\rho_{ij}}{\partial t}=\sum_{i',j'\in \mathcal{N}_{ij}} \frac{e^{\beta A_{ij}}}{Z_{i'j'}}\rho^t_{i'j'}-\rho_{ij}^t
\end{equation}
The next step is expanding the RHS. Note that the effective 'neighborhood' that we need to consider is actually a $5\times5$ neighborhood. Although $\mathcal{N}(i,j)$ is $3\times3$, we are now summing $Z_{i'j'}$ over the neighborhood, and each $Z_{i'j'}$ depends on $\mathcal{N}_{i'j'}$, such that the contributions to  $\frac{\partial\rho_{ij}}{\partial t}$ come from a $5\times 5$ neighborhood effectively.

To start expanding the LHS, we simply assume that adjacent cells are a distance $\Delta x$ apart from each other. Further, we assume that both $\rho_{ij}$ and $A_{ij}$ are smooth. Then, we start Taylor expanding into increasing $\Delta x$ orders. Since at $\beta=0$ we expect to find diffusion (see (\ref{eq:true_eqs})), we should expect the RHS's first non-vanishing term to be of order $O(\Delta x^2)$. Indeed, that is exactly what happens, and we are left with the following approximation to the LHS, after performing the expansion with the help of Mathematica :

\begin{equation}
    \Delta t \partial_t\rho = \frac{\Delta x^2}{3} \left(\partial_x^2\rho+\partial_y^2\rho-2\beta\partial_x\left(\rho\partial_x A\right)-2\beta\partial_y\left(\rho\partial_y A\right)\right)+O(\Delta x^3)
    \label{eq:taylor_expanded}
\end{equation}
Where we can see that it corresponds to (\ref{eq:continuousunnormalized}). The only difference is the $O(\Delta x^4)$ term : indeed, the $O(\Delta x^3)$ vanishes, and this comes from the fact that the discretization is parity-invariant both in the $y$ and $x$ direction, so odd terms vanish.

Now, given (\ref{eq:taylor_expanded}), we know what rescaling we should apply to the LHS and RHS such that they have a non-trivial continuous limit. In this case, we need to rescale the LHS by $\Delta t$, and the RHS by $\Delta x^2$. Let's go back to (\ref{eq:arrangerhocomp}) and do that :

\begin{equations}
        &\frac{\rho_{ij}^{t+\Delta t}-\rho_{ij}^t}{\Delta t} = \frac{3}{\Delta x^2} \left( \sum_{i',j'\in \mathcal{N}_{ij}} \frac{e^{\beta A_{ij}}}{Z_{i'j'}}\rho^t_{i'j'} - \rho^t_{ij}\right) \label{eq:true_eq_v2}\\
        &\rho^{t+\Delta t}_{ij}=\rho_{ij}^t(1-\frac{3\Delta t}{\Delta x^2})+\frac{3\Delta t}{\Delta x^2}\sum_{i',j'\in \mathcal{N}_{ij}} \frac{e^{\beta A_{ij}}}{Z_{i'j'}}\rho^t_{i'j'}
\end{equations}
And we recover the equations in (\ref{eq:true_eqs}). Note that we divided by $\frac{\Delta x^2}{3}$ on the RHS instead of $\Delta x^2$; this is simply a choice, we choose a normalization such that at $\beta=0$ we have diffusion with diffusion coefficient $D=1$.

We can see from (\ref{eq:true_eq_v2}) that intuitively, we require $\frac{3\Delta t}{\Delta x^2}<1$ for stability. Indeed, in the limit $\beta\rightarrow 0$  we can show this sufficient explicitely  by means of a Von Neumann analysis. We can compare to the usual scheme (where diffusion is implemented with a $3\times  3$ uniform stencil), which requires $\frac{4\Delta t}{\Delta x^2}<1$, and is thus less stable. Intuitively, this is simply because in our case, even at $\beta=0$, we are incorporating information from a larger $5\times 5$ neighborhood which tends to stabilize the dynamics.

Note that if $A_{ij}$ depends on $\rho_{ij}$, the condition $\frac{3\Delta t}{\Delta x^2}\leq 1$ might not be sufficient. Although the dynamics will never explode in this case, they could develop artifacts. Empirically, for MaCELenia we have had no problems, although for example adding a 'regularization term' of the form $A \rightarrow A-\frac{c_{reg}} \rho$ (as in Flow-Lenia \cite{plantecFlowLeniaOpenendedEvolution2023a}) requires slightly lower values of $\frac{3\Delta t}{\Delta x^2}$, about $0.85$.

\section{Random Fourier Functions} \label{app:randomfourier}
In this appendix, we describe in a bit more detail how we draw 'random functions' by sampling Fourier coefficients. The main idea is very simple; we would like to draw random \emph{smooth} functions, and to do that we sample Fourier coefficients and then reconstruct the function from them. In this way, the smoothness is automatic.

The basic equation underlying this method is (\ref{eq:arbifunctions}), which is simply the Fourier decomposition for functions $f$ of period $P$.
\begin{equations}
\text{sample }a_k, b_k \in \mathbb R \rightarrow f(x) &= \sum_{k=1} b_k \sin(2\pi kx/P)\\&+\sum_{k=0}a_k \cos(2\pi kx/P), \,\, x\in [0,P] \label{eq:arbifunctions}
\end{equations}

When generating functions in this way, we introduced several parameters that allow us to control a bit more the functions that we sample : 
\begin{itemize}
    \item \textbf{Harmonics}: When generating, we can choose how many Harmonics, and which harmonics to include. We usually limit ourselves to $\leq 10$ harmonics.
    \item \textbf{Range}: This parameter is a tuple $[a,b]$, which specifies the x-range of the function. To force the function in this range, it suffices to set $P=a-b$ and $x\rightarrow x-b$ in (\ref{eq:arbifunctions}).
    \item \textbf{Rescale}: In some cases, we want a function with y-values in a specific range. To force this, we first sample the Fourier coefficient normally, then rescale the function linearly to lie in the desired range.
    \item \textbf{Clip}: This parameter allows us to clamp the function to restrict it to a defined range of y-values. Similar in use to rescale, but clips the values instead.
    \item \textbf{Decay}: We have the option to decay exponentially higher harmonic, to generate functions with multiple scale variations. We simply multiply by an additional $e^{-d}$ factor for each harmonic we add, where $d$ is the decay strength.
\end{itemize}

Generally, for MaCELenia, we find that generating random Fourier kernels provides nice results, if pair with a clipping of low values, that gives rise to the usual 'bumps'. It is more diverse than using Gaussian bumps, but also harder to search.

We also experimented using random Growth functions. In our experimentation, this did not produce more diverse behaviours, at least qualitatively. Additionally, the computational cost of evaluating the Growth function slows down the dynamics as soon as we have more than $4$ harmonics. This is not a problem for the kernel, as it is computed once, then stored.

To use growth functions with more harmonics in real-time, a possible solution would be to approximate it using piecewise linear pieces (or any other approximation which can be retrieved quickly), and store this information, allowing the computation of the growth to be a simple lookup.

Overall, it would be interesting to run Quality-Diversity search \cite{faldor2024artificialopenendedevolutionlenia} or another open-ended search algorithm, to understand whether the increased expressivity of the random Fourier functions allows for more diverse behaviours.

\section{Details on random sampling in MaCELenia}\label{app:random_sampling}
All experiments are done using a kernel size of $31$ pixels, which corresponds to $\Delta x=2/31\approx 0.06$, assuming the kernel is size $2$, and sets the x-scale.

In Lenia, the free parameters are those used to define the growth functions, as well as the convolution kernels which are used. MaCELenia only adds two additional parameters, which is the inverse temperature $\beta$ and the relative discretization of $\Delta t$ and $\Delta x$. In what follows, we describe the parameters in detail, and table \ref{tab:randoparamtable} summarizes how they were sampled.

First, for the growth function, we need a pair $(\mu,\sigma)$ for the Gaussian bumps, one pair for each channel-channel interaction, for a total of 9 pairs. There are also 9 weights $w$, which determine the relative effect of the growths, see (\ref{eq:growthlenia}), where they are denoted $W_{c\tilde{c}}$.

For the kernels, we have again 9, one for each channel-channel interaction. Each kernel is composed of three Gaussian bumps, centered at $\mu_k$ with standard deviation $\sigma_k$. Additionally, each bump has a relative size (amplitude) given by the numbers $b_k$.

In the table \ref{tab:randoparamtable}, when we indicate a range it means the parameter is sampled uniformly, otherwise we specify the random variable we used for the sampling.
\begin{table}[h]
\centering
\begin{tabular}{ll|ll}
\toprule
\multicolumn{2}{l|}{\textbf{Growth fields}} & \multicolumn{2}{l}{\textbf{Kernels}} \\
\midrule
$\mu$ & $ [0.0, 0.7]$ & $\mu_k$ & $0.5+0.2\mathcal{N}(0,1)$ \\
$\sigma$ & $\left[0.0, 0.9 \frac{\mu}{2\ln 2} \right]$ & $\sigma_k$ & $\text{max}\left[0.05(1+0.3\mathcal{N}(0,1)),0.005\right]$ \\
$w$ & $ [0, 1]$ & $b_k$ & $ [0, 1]$ \\
\midrule
\multicolumn{4}{l}{\textbf{MaCE}} \\
\midrule
\multicolumn{2}{l}{$\beta = 8$} & \multicolumn{2}{l}{$\dfrac{3\Delta t}{\Delta x^2} = 1$} \\
\bottomrule
\end{tabular}
\caption{'Smart' sampling}
\label{tab:randoparamtable}
\end{table}

The main feature of this sampling is the sampling of $\sigma$, which is dependent on the sampled $\mu$. It was derived in \cite{papadopoulos2024lookingcomplexityphaseboundaries}, and is supposed to produce Lenia parameters which are close to the 'phase-transition region' between death and explosion. As explained in the main text, we find this sampling to generate lots of highly mobile rules, with several solitons, although the RGB colors tend to separate.

The other sampling experiments were done with the same prior, except now we choose $\sigma \in [0.0,0.3]$, uniformly. As explained in the text, this results in less mobile rules on average, but we find richer color mixing. 

\section{Comparison to FlowLenia}\label{app:compare_flow}
\begin{figure}
    \centering
    \includegraphics[width=.9\linewidth]{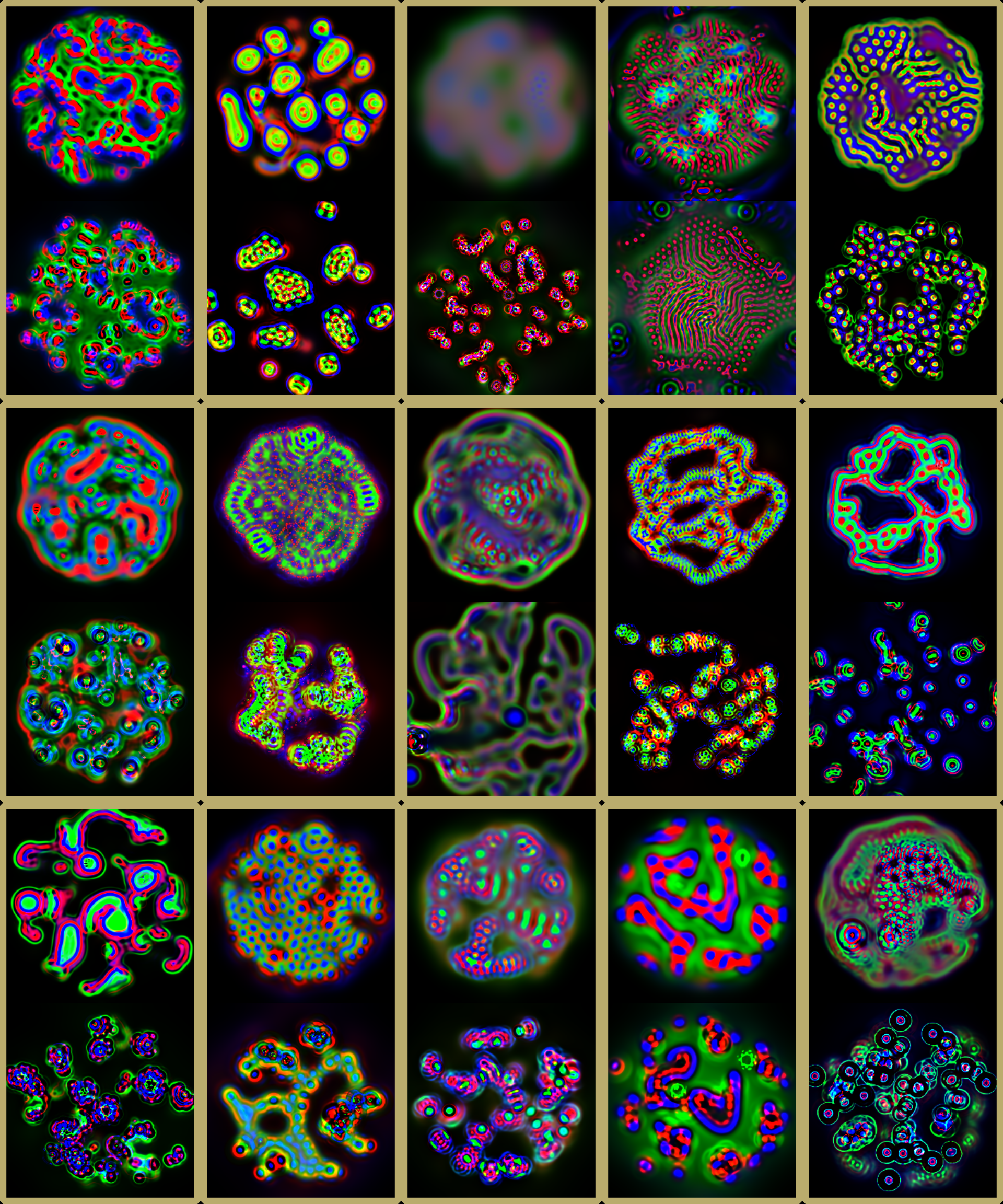}
    \caption{Snapshots of FlowLenia vs MaCELenia after 1500 steps, Lenia parameters generated with the 'random' sampling. In the yellow rectangles, top image is FlowLenia, bottom is MaCELenia.}
    \label{fig:macevsflowrandom}
\end{figure}

\begin{figure}
    \centering
    \includegraphics[width=0.9\linewidth]{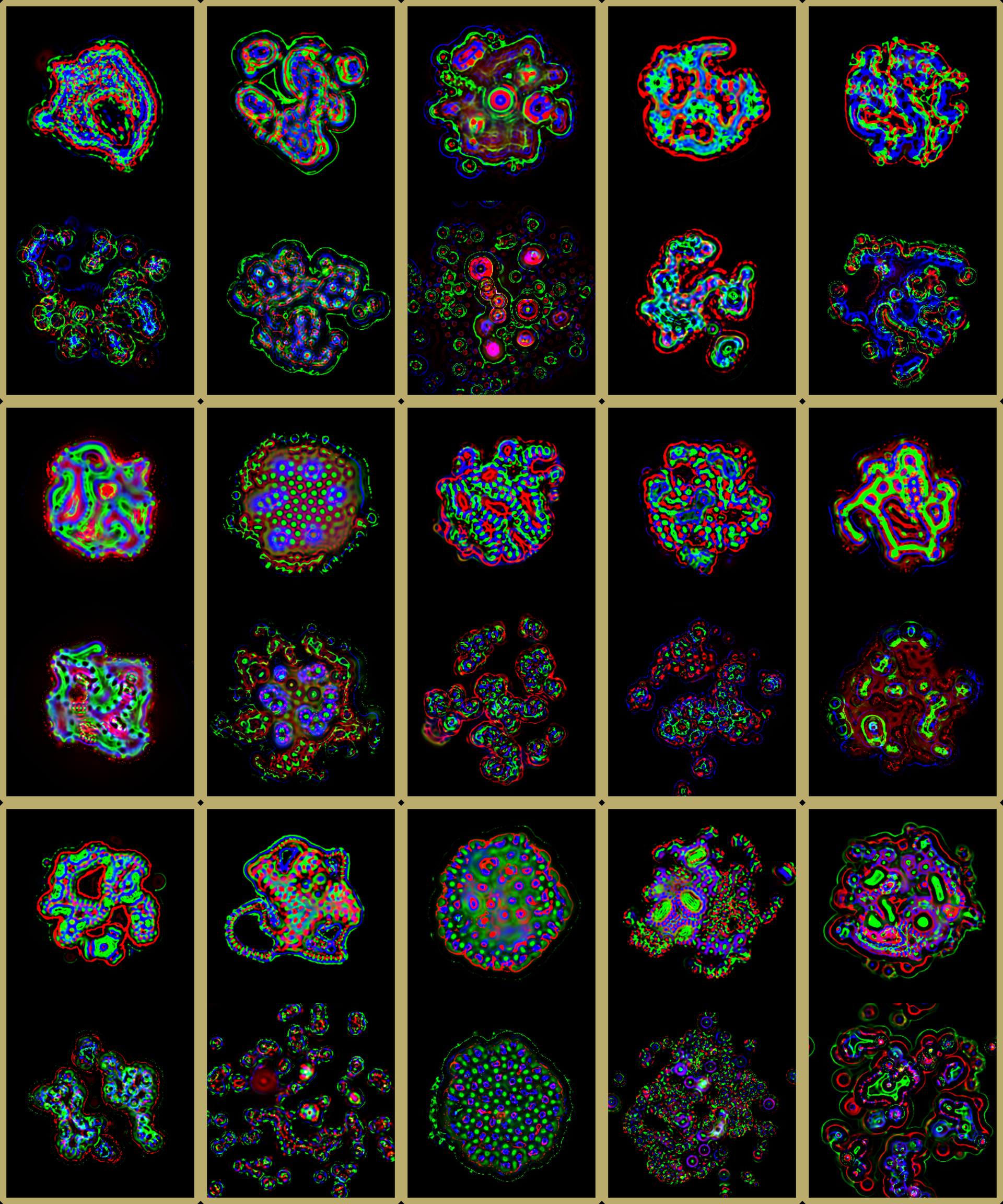}
    \caption{Snapshots of FlowLenia vs MaCELenia after 1500 steps, Lenia parameters generated with the 'smart' sampling. In the yellow rectangles, top image is FlowLenia, bottom is MaCELenia.}
    \label{fig:macevsflowdefault}
\end{figure}

In this section we provide some snapshot of FlowLenia (as re-implemented by us) and MaCELenia, with the same exact parameters for Lenia. We set MaCELenia to $\beta=7$, and we tune manually the free parameters of FlowLenia to promote maximal movement (in practice, we have $dd=2$, $\theta_A=3.5$, $s=0.6$, $dt=0.13$, $n=2$). Then, we initialize a state, and evolve it with the same number of frames, and provide snapshots here. Fig. \ref{fig:macevsflowrandom} displays the final states for parameters sampled with 'random' sampling, whereas Fig. \ref{fig:macevsflowdefault} displays the final state for parameters sampled with 'smart' sampling.

Both for 'random' and 'smart' sampling, MaCELenia generally produce dynamics that tend to 'separate' more, in thes sense that the initial blob cuts itself into smaller pieces, often producing moving solitons. The final snapshots look qualitatively similar (in terms of color distribution, and turing patterns), which is to be expected as in both cases, the computation of the affinities is done exactly in the same way. In terms of dynamics, MaCELenia dynamics are much 'faster', in the sense that the movement/flow of matter appears much faster, given the same number of update frames. This is probably not due to 'differing' $\Delta t$'s, although they are hard to compare. For $k_size=31$, the effective $\Delta t$ that we are using (with the update as in (\ref{eq:rhocomp})) is $\Delta t \sim 0.001$. In contrast, the $\Delta t$ in Flow Lenia is much higher, $\Delta t\sim 0.1$, so we would expect it to have faster dynamics. Several explanations are possible  :
\begin{itemize}
    \item Because of the regularization flow term, and the $s$ parameter in the flow, affinity gradients are generally lower, slowing down the flow
    \item Since the continuous limit is not formally defined in Flow Lenia, it is possible that the effective $\Delta t$ is different from the one appearing in the Reintegration Tracking
    \item The $\beta$ parameter in MaCE acts as a $\Delta t$ multiplier. $\beta =10$ increases the Flow values ten-fold, which speeds up the dynamics (in addition to altering them).
\end{itemize}

FlowLenia displays better color mixing than MaCELenia, although this is mostly noticeable for the 'smart' sampling regime, which tends to cause more color separation. This may be due specifically to the 'repulsive' parameter that is added in FlowLenia, which essentially adds 'negative affinity' in regions of high concentrations. This prevents very high concentration, and as such colors tend to spread better, and hence mix. Note that since this 'repelling force' is simply a modification of the affinity term, it can very well be included in the MaCE update, simply by modifying the affinity computation\footnote{Note that the dynamics will still not be identical, as the MaCE update always involves a diffusion term. It is unclear whether the '$s$' parameter in Flow Lenia amounts to a diffusion term, or something else, in the continuous limit.}. In our brief tests, we find that this renders the system a bit less stable, and $\frac{3\Delta t}{\Delta x^2}\leq1$ does not ensure stability anymore. However, slightly lowering this value below one restores stability. We leave for future work to explore the variants of MaCELenia that can be obtained by modifying its update rule.

We also include full video comparison of the dynamics in the video page, that can be found at \href{https://etimush.github.io/MaCE-Videos/}{this link}.
\end{document}